\documentclass[10pt,twocolumn]{article} 
\pdfoutput=1
\usepackage{simpleConference}
\usepackage{times}
\usepackage{graphicx}
\usepackage{amssymb}
\usepackage[bookmarks=false]{hyperref}

\begin{document}

\title{Taxonomic Classification of IoT Smart Home Voice Control}
 
\author{Mary~Hewitt, Hamish~Cunningham \\ \\
Department of Computer Science, University of Sheffield, UK \\
{\tt \{mjhewitt1, h.cunningham\}@sheffield.ac.uk} \\
{\tt hamish@gate.ac.uk}
\thanks{Centre for Doctoral Training in Speech and Language Technologies funded by UK Research and Innovation (UKRI) grant number EP/S023062/1}.}

\maketitle

\begin{abstract}
Voice control in the smart home is commonplace, enabling the convenient control of smart home Internet of Things hubs, gateways and devices, along with information seeking dialogues. Cloud-based voice assistants are used to facilitate the interaction, yet privacy concerns surround the cloud analysis of data. To what extent can voice control be performed using purely local computation, to ensure user data remains private? In this paper we present a taxonomy of the voice control technologies present in commercial smart home systems. We first review literature on the topic, and summarise relevant work categorising IoT devices and voice control in the home. The taxonomic classification of these entities is then presented, and we analyse our findings. Following on, we turn to academic efforts in implementing and evaluating voice-controlled smart home set-ups, and we then discuss open-source libraries and devices that are applicable to the design of a privacy-preserving voice assistant for smart homes and the IoT. Towards the end, we consider additional technologies and methods that could support a cloud-free voice assistant, and conclude the work. 
\end{abstract}

\section{Introduction}
The Internet of Things (IoT) describes a vision where objects become part of the Internet, and expand it in such a way that our digital and physical worlds are fused together \cite{Coetzee2011-jp}. At a technical level this vision is actualised using networked microcontrollers, sensors and actuators \cite{Cunningham}. These devices are consistently subject to downward pressure on cost and energy consumption, similarly to embedded electronics and distinct from general purpose computers (e.g. desktops, smartphones).

Smart homes are understood to consist of a network of interconnected devices and sensors, that seamlessly communicate with each other, and can be controlled by the user in a convenient way \cite{Gram-Hanssen2018-ik}. The primary user benefits of smart homes include saving energy, enabling comfier, healthier living environments and ensuring home safety and security \cite{Nicholls2020-eq}. These days, the IoT is viewed as an important technology towards improving living environments and quality of life \cite{Wang2021-zn}. The number of publications on IoT-based smart homes was seen to grow significantly between 2015 and 2019 \cite{Choi2021-ju}, reflecting the increase in research interest in this field.  

Advances in Automatic Speech Recognition (ASR) technologies have given rise to voice-controlled smart homes \cite{Poongothai2018-pf}, and the market is now populated with devices to build these, e.g. Samsung's SmartThings, Amazon’s Alexa, Apple’s HomeKit and Google’s Home Assistant. The uptake of voice-controlled smart home devices has been slowed due to privacy and trust concerns that stem from the product’s reliance on cloud-based data analysis \cite{Lau2018-vl, Brush2011-rb}. Secrecy surrounds corporate practices, meaning companies can capture user conversation and preferences without them being aware of what, how or why data is recorded \cite{Nicholls2020-eq}. Consumers have expressed specific concerns regarding lack of control over their data, audio and video access, household profiling, government access, and data breaches \cite{Haney2020-fc} \cite{Marikyan2019-pz}. 

Recent events draw attention to smart home privacy and security concerns. It was reported that Amazon’s Ring doorbell can give data to the police without needing your knowledge or consent \cite{Morrison2022-in}, and fraudulent legal requests have caused some technology companies to provide sensitive information about their customers \cite{Turton2022-oo}. Additionally, a security analysis of the Samsung SmartThings framework found it was possible to steal lock pin-codes and cause fake fire alarms through exploiting design flaws and vulnerabilities of over-privileged third-party apps \cite{Fernandes2016-uc}. 75\% of people agree there is reason for concern about their data being used by other organisations without their permission, and security concerns deter almost a third of people who do not own smart devices from buying one. \footnote{https://www.internetsociety.org/wp-content/uploads/2019/05/\\CI\_IS\_Joint\_Report\-EN.pdf} The use of local data processing and authentication methods have been suggested to protect users rights and abide by GDPR principles \cite{Hernandez_Acosta2022-ux}.

Edge computing has emerged as a paradigm in which computing and storage resources are placed in close proximity to end users on devices or sensors \cite{Satyanarayanan2017-xv}. In contrast with cloud-based systems that suffer from power hungry components, high latency and privacy and security concerns \cite{Pinto2020-sc}, edge computing brings advantages in terms of energy savings, bandwidth savings, privacy protection, reliability, low-cost components and low-latency \cite{Wang2022-tl} \cite{Wang2020-ge}. Edge computing can ensure the security and privacy of a network \cite{Ding2022-lf} and it is predicted that in the next decade most speech recognition will happen on the device or at the edge \cite{Hannun2021-dn}.

The paper arises from an investigation into the feasibility of implementing voice control at the edge, as we work towards a more ideal smart home than those commercially provided. As a first step in answering this question, we examine commercial smart home systems and present a taxonomic classification of the current technologies enabling voice-control in the smart home. The paper begins detailing background literature in the field, we then present our taxonomy and turn to examine academic efforts in implementing and evaluating voice-controlled smart homes. In line with our aims, we consider the existing landscape of offline speech recognition tools and implementations, and also discuss further methods that could help towards a \textit{privacy-preserving} voice-controlled smart home. 

\begin{itemize}
  \item \textbf{Section 2} gives a high-level overview of typical smart home hardware architectures, and outlines the technologies that comprise voice assistant software. 
  \item \textbf{Section 3} considers previous studies that categorise the devices and technologies involved in voice-controlled smart home setups in order to inform the design of our taxonomy.
  \item \textbf{Section 4} presents our taxonomy of commercially available voice-controlled smart home technologies and devices, and offers some discussion and analysis of the categories defined.
  \item \textbf{Section 5} considers academic-based smart home implementations with consideration for the devices and voice assistant technologies employed, as well as evaluation methods that assess the performance of these.
  \item In \textbf{Section 6} we identify and compare currently available speech recognition and voice assistant systems that have been or could be used for voice-control in a smart home set-up. Tools and techniques used are discussed, as well as evaluation methods used.  
  \item \textbf{Section 7} envisions the design of a better voice-controlled smart home. We consider the use of cheaper, programmable devices, authentication methods, and model personalisation.
\end{itemize}

\section{Background}

\subsection{Smart Home IoT Architectures}
At a technical level, we simplify the IoT to center on the use of networked microcontroller devices, that combine in architectures that typically include microprocesser-based hubs or gateways, and a cloud-based server side \cite{Cunningham}. We define smart homes to be systems that combine networked devices that seek to replace or augment the control mechanisms that have matured gradually over previous decades: TV remotes, central heating thermostats or washing machine programmers.

Smart home systems can either be centralised or distributed. A centralised gateway architecture is generally seen to be optimal for supporting multiple resource-constrained devices \cite{Lin2016-th}, where the gateway serves to coordinate devices, and connect the local infrastructure to the internet \cite{Samuel2016-lr}. Upper and lower parts of a smart home system have been defined in \cite{Wang2013-cn}, with the upper part consisting of a wireless router, computers, tablets, and the lower part consisting of switch modules, data collectors, plus a smart central controller to connect the parts together. A typical gateway architecture for IoT devices is described in \cite{Kruger2014-yw}. 

\begin{figure}
\centering
\includegraphics[width=1.8in]{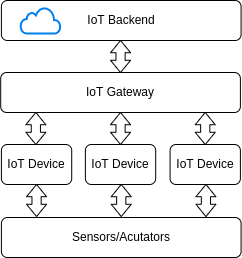}
\caption{Smart Home IoT architecture}
\label{sys_diagram}
\end{figure}

Market leaders usually provide a central hub (or gateway) for smart homes, where compatible smart devices can be purchased to connect with the hub and establish a smart home network, e.g. in product descriptions you can often see “works with X” \cite{Serrenho2019-bk}. Figure \ref{sys_diagram} presents an architecture diagram of a gateway-based IoT smart home system that makes use of cloud-processing. In the case of non-cloud systems, the processing and storage handled by the IoT backend (cloud) is usually placed at the gateway or IoT devices. Sensors and actuators are commonly attached to edge devices, where sensors provide continuous data streams about an environment \cite{Sudharsan2022-yf}, and actuators act on the environment in some way (e.g. switch).

Wireless technologies allow the flexibility to add and remove components to the smart home network, enabling scalability and expansion \cite{Viani2013-uj}. Communication types commonly used in smart homes include Wi-Fi, Infrared, Radio Frequency (RF), and Bluetooth \cite{Katuk2018-zj}, plus Global System Mobile (GSM), Z-Wave, ZigBee, and wired connections (e.g. Ethernet) can also be used \cite{Arriany2016-cv}.

\subsection{Voice Control Technologies}
Voice control entails computational transcription of the spoken word, and the interpretation of user intentions for device control or information seeking. Voice-controlled systems are described in terms of three modules in \cite{Mishakova2019-qd}: Automatic Speech Recognition (ASR), Natural Language Understanding (NLU) and Decision Making. Alternatively, \cite{Vacher2015-dv} define voice-controlled dialogue systems to be composed of five stages: Voice Activity Detection (VAD), ASR, NLU, a decision stage, and a communication stage. 

\begin{figure}
\centering
\includegraphics[width=1.8in]{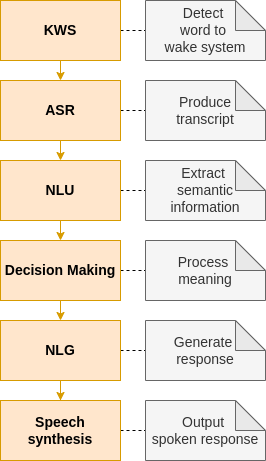}
\caption{Voice Assistant system components}
\label{vasystem}
\end{figure}

We define voice control in the home to entail keyword spotting (KWS) for system entry, ASR for transcribing user utterances, and NLU for interpreting the action specified in the utterance. Following \cite{Huang2015-db}, we define voice assistants (VAs) to build on these core functionalities, and make use of a dialogue manager, natural language generator (NLG) and speech synthesis to enable the two-way interaction seen in dialogue systems. We present the relationship between our defined modules comprising voice assistants in Figure \ref{vasystem}

Spoken Language Understanding (SLU) refers specifically to the task of inferring the meaning or intent of a spoken utterance \cite{Lugosch2019-bg}. ASR and NLU modules comprise a conventional SLU system. More recently end-to-end SLU systems have gained popularity, where a single model is used to map speech input directly to user intent, without the intermediary step of producing a transcript. By jointly optimising the ASR and NLU components, cascading errors are reduced which helps training and gives the technique an advantage \cite{Desot2022-ub}.

There exist a wide range of additional speech processing tasks that voice assistants have been seen to use. For example, smart homes that have multiple voice-enabled devices use device arbitration, speech enhancement, and speech localization models to improve the performance and user experience \cite{Ciccarelli2022-uh}. We primarily focus on ASR and NLU components in examining voice assistants in this paper, since these are integral to voice-control in the smart home. All voice assistants use ASR for recognising what the user has said, and NLU is necessary for finding meaning in natural language commands.

\subsection{Smart Homes, Voice-Control, and the IoT}
\begin{figure}[!t]
\centering
\includegraphics[width=2.8in]{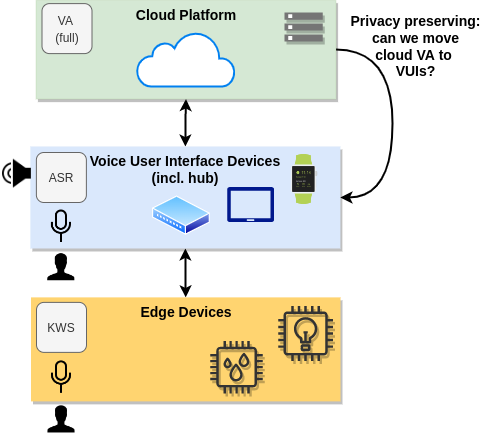}
\caption{Voice-control in the smart home}
\label{homearch}
\end{figure}

Figure \ref{homearch} shows how the IoT, smart home and voice control technologies operate in conjunction. Our diagram is informed by commercial smart home and IoT architectures. We introduce each component and the voice-control capabilities of each: 

\begin{itemize}
    \item Cloud-based computation is the current predominant method of processing speech data, and is capable of running complex algorithms to allow for voice assistant (VA) interaction. 
    \item Voice User Interface (VUI) devices are equipped with microphones, and both collect and transmit user commands to the cloud for processing, as well as  co-ordinate the corresponding action execution via communication with edge devices. VUI devices tend to be based on micro{\em processors} which can handle some VA functionality such as ASR. 
    \item The edge devices are the distinctive category in the IoT, distinguishing the field from general purpose computing. They are usually low power, MCU based devices, that sense or act on the home environment. The tightly constrained nature of edge devices limits their speech processing capabilities to simple KWS algorithms at present.
\end{itemize}

\section{Existing Categorisations}

\subsection{System and Device Categorisations}
The importance of having clear terminology relating to IoT devices has been stressed \cite{Haller2010-yx}, but there exist a variety of diverse smart home devices that are somewhat difficult to categorise. We analyse previous efforts to classify and categorise aspects of smart home systems and devices, in order to inform the design of our taxonomy that is intended to cover a wide range of currently available smart home devices. 

\cite{Koreshoff2013-eg} consider human interaction with the IoT, with the aim of comparing commercial products with academic research efforts in the field to observe trends, gaps, differences and common areas of effort. These aims closely mirror our own, making the paper highly relevant to our work. In the study, IoT-based products are classed as either person-centric if they gather data about the human body, or home-centric if they gather data about the environment. Devices in each category are detailed in terms of their input and output sensor types and means of user interaction. Insights about current trends can be made using the device-level detail, e.g. person-centric devices most commonly contain accelerometers, therefore we will follow aspects of the work in our taxonomy design. 

\cite{Alam2012-zl} categorise smart home devices, with consideration for smart home services offered (comfort, healthcare, security), as well as the devices that enable these services. Devices are categorised as either a sensor (acquire data from the environment), a physiological device (monitor health conditions), or a multimedia device (provide an interface between the system and the user). Example devices for each category are given, and we look to a similar approach where we characterise devices by the hardware-level abilities of the device, in combination with the type of service it provides the user.

\cite{Katuk2018-zj} classify devices based on the room each resides in, with attention placed on product manufacturers and smart features of devices, e.g. LED lights are manufactured by Philips and you can select preferred lighting modes. Identifying the functionality of devices and leading manufacturers of smart home devices is useful. Categorising devices based on room appears restrictive and not that informative, since homes can have a range of layouts.

\cite{Kumar2019-mq} analyse the use of IoT devices in home networks, and define fourteen categories of smart home device: Computer, Network node (e.g. home router), Mobile device (e.g. iPhone or Android), Wearable (e.g. Apple Watch), Game console (e.g. XBox), Home automation (e.g. Nest Thermostat), Storage (e.g. home NAS), Surveillance (e.g., IP camera), Work appliance (e.g. printer), Home appliance (e.g. smart fridge), Generic IoT (e.g. toothbrush), Vehicle (e.g. Tesla), Media/TV (e.g. Roku), Home Voice Assistant (e.g. Alexa). The work focuses on the popularity of the smart home device categories, with discussion on device vendors. A wide range of devices are seen to be used, yet there is limited discussion on the characteristics of these.

User benefits are a common way to categorise smart home systems, which does not provide much insight into the hardware and specific capabilities of devices. For example, \cite{Holroyd2010-gr} define three classes of smart home user benefits: energy saving, support for elderly or disabled, security and safety. \cite{De_Silva2012-tm} similarly identify four applications of smart home devices: healthcare, better life, security, energy efficiency. \cite{Arriany2016-cv} categorise smart home applications as convenience and entertainment, safety and security, energy savings, and healthcare. While useful to consider, there is limited scope for insights to be made using such categorisation methods. 

Categorising device components based on their role in the smart home network is also commonly seen. \cite{Suh2008-zt} define sensors to gather home environment data, actuators to control home devices, control to components manage the actuators, decision components to select services based on sensor data, and service components to be the software that provide the user benefits. Similarly \cite{P_R_Filho2018-cx}, categorise devices in a smart home network as sensor nodes, decider nodes, actuator nodes and sink nodes. Likewise, \cite{Sun2013-gu} define smart homes to have sensing agents (e.g. temperature sensor), action agents (e.g. door lock), administration and decision agents (e.g. smart speaker), and database agents (e.g. knowledge bases). Furthermore, \cite{Gunge2016-rc} define a home automation system to have a User Interface (UI), mode of transmission (wired or wireless), a central controller (a hardware interface) and electronic devices (compatible with transmission mode and connected to the central controller). Studies like these, that define classes of hardware devices based on their functional role and capabilities in the home network, are useful towards the design of our taxonomy. There can be crossover between categories however, that are not considered in these works. 

Recognising that smart home device classifications can fail to accommodate devices with multiple functionalities, \cite{Lopez2011-fz} propose the ISADN specification, where devices are categorised as having Identity, Sensors, Actuators, Network connectivity, and Decision making abilities. The specification is intended to help describe the characteristic functionality of smart objects, rather than pose a constrained view of smart devices. The terminology helps towards differentiating and describing the nature of hardware devices based on their abilities.

Distinct from other works, \cite{Sturgess2018-pv} reduce smart home devices to their data-collecting capabilities, so that they can assess the privacy risk of the system based on the information the user exposes. We will also examine the type of sensors found on each device in our taxonomy in order to place awareness and focus on the data types that can be captured by smart home devices.

\subsection{Voice Assistant Categorisations}
Voice assistants can be distinguished as being manually activated, speech activated, or always on \cite{Hernandez_Acosta2022-ux}. For example, Alexa (used in Amazon Echo) is a cloud-based voice service that is always on, and therefore records all voice activity in the home even when it is not activated \cite{Venkatraman2021-ly}. Voice assistant systems can also be characterised by the types of speech act it can understand and respond to, e.g. speech acts can inquire about information, control devices and request services \cite{Huang2015-db}. Response styles of popular commercial virtual assistants have been categorised as either minimal, keyword and full sentence, by analysing responses to frequently used queries and commands \cite{Haas2022-ej}. 

ASR is a core component of voice assistants. Criteria for classifying ASR systems more specifically, considers isolated vs. continuous speech, speaker dependent vs. independent models, dictation vs. spontaneous speech styles and vocabulary size \cite{Peinl2020-fv}. \cite{Arriany2016-cv} classify ASR systems as speaker-based or word-based. Word-based systems are categorised by how the speaker says the words in the sentence, e.g. discrete words or connected words (continuous speech). Speaker-based systems can either be speaker dependent or speaker independent. Speaker dependent systems use template matching and are trained on certain voices or words before use, and speaker independent systems perform feature analysis to analyse the input voice.

\section{Taxonomy}

\subsection{Taxonomy Overview}
We present a taxonomy of the components of voice-controlled smart home systems, from both a software and hardware perspective by studying commercially available systems. An overview of the taxonomy design is first introduced, and then we draw attention to specific parts of the taxonomy.

The high level view of the taxonomy is informed by identifying core components in a commercial smart home system architecture. We define three device categories that make-up a voice-enabled smart home system from a user interaction perspective: Voice Assistant, Voice User Interface, Edge Device. 

\begin{figure}[!t]
\centering
\includegraphics[width=2.3in]{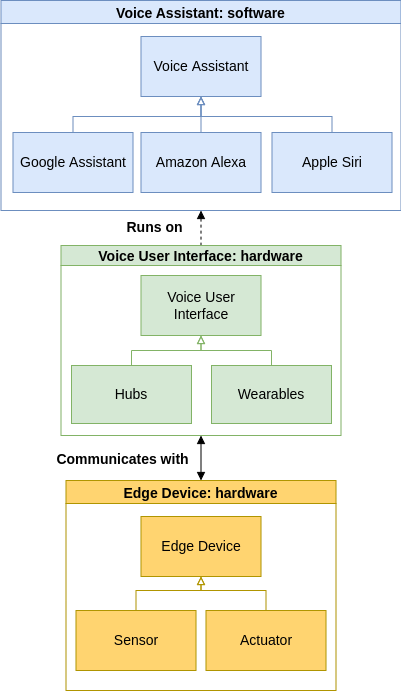}
\caption{Taxonomy}
\label{fig:taxonomy}
\end{figure}

On the software side, we consider the voice assistants used in voice-controlled smart homes to process and respond to spoken commands. Specifically, we consider Amazon Alexa, Google Assistant and Apple Siri since they are widely used and popular commercial solutions to smart home voice-control. On the hardware side, we consider VUI’s that receive voice input and have abilities to communicate and control edge devices. Edge devices are low power devices, typical of the IoT, and designed specifically to sense and/or act on aspects of smart home environments. The taxonomy overview is depicted in Figure \ref{fig:taxonomy}, where we show the relationship between categories of smart home technology. Hubs and wearables are defined as types of VUI, and sensors and actuators are types of edge device.

Crossover between hardware device categories can occur, therefore we distinguish the categories. Hubs are generally home-centric gateway devices, requiring more power than wearable devices that reside on the person. VUI devices usually contain CPUs and are more power hungry than typically low-power MCU-containing edge devices \footnote{The core distinction is between microcontroller (MCU) and microprocessor (which we're abbreviating as CPU here). Strictly speaking, the latter includes more than just a CPU, but it is relatively common to refer to microprocessors in this way \cite{Schlett}.}. In some cases, a VUI could also be considered an edge device, (e.g. a smart speaker). The differentiation between a VUI and a edge device arises from the general purpose capabilities a voice assistant provides a VUI device (e.g. setting timers, making calls, controlling other devices). Edge devices are usually special purpose, but can have both sensing and actuation abilities, therefore we generally characterise these devices by classifying each based on its predominant function. Outside of the scope of the taxonomy lie general purpose computing and networking devices that are not exclusive to smart homes, such as routers, mobile devices, games consoles and computers. Third-party services that provide additional functionality to commercial smart homes are also not considered. 

We move on to take a specific look at the software and hardware components we have outlined in our taxonomy. Each of the three top-level categories is discussed individually across the following three subsections.

\subsection{Voice Assistants}
Three classes of voice assistant software are defined, as shown in the taxonomy overview (Figure \ref{fig:taxonomy}): Google Assistant, Amazon Alexa and Apple Siri. The voice assistants from these three massive corporations are selected for inclusion given their significance, popularity and wide-spread use in the realm of commercial voice-enabled smart home systems. We seek to identify the similarities and differences between each by examining their capabilities, the available keywords for waking the device ({\em wakewords}), the accepted command types and the response behaviours. 

Our findings are detailed in Table \ref{tab:VA}, and reflect the similarities between commercially available voice assistants. While extensive, the functionality is consistent amongst each, with minimal variation in the accepted commands and types of response styles. Third-party apps are available for each to extend capabilities, and as expected, none are open-source and all use cloud-based processing. Combined, these characteristics point up the lack of transparency, flexibility and diversity in commercially available voice assistants. From this perspective, we can conclude that there is space for an alternative voice assistant that can meet users' needs without compromising their privacy. Our analysis of voice assistant functionalities supports research into how these could be replicated using on-device processing methods. In Section 6, we discuss currently available methods to achieve this. 

\begin{table*}[t]\footnotesize
\centering
\begin{tabular}{|p{1.3cm}|p{9cm}|p{1.5cm}|p{1.5cm}|p{2cm}|}
\hline
{\bf Name} & {\bf Capabilities} & {\bf Wakeword} & {\bf Command input type} & {\bf Response behaviour} \\ [0.5ex]
\hline \hline
Google Assistant & Control smart home devices; play and control media; search and retrieve information; manage alarms, timers, lists, calendars, tasks; play games; purchase items; make calls and announcements & Hey Google, Ok Google & Full sentence & Full sentence ('brief' mode available) \\ \hline
Amazon Alexa & Control smart home devices; play and control media; search and retrieve information; manage alarms, timers, lists, calendars, notes; play games; purchase items; make calls and announcements & Alexa, Amazon, Echo, Computer & Full sentence & Full sentence ('brief' mode available) \\ \hline
Apple Siri & Control smart home devices; play and control media; search and retrieve information; make calls, texts, announcements, payments; manage alarms, timers, lists, calendars, notes, reminders & Hey Siri & Full sentence & Full sentence \\ [1ex]
\hline
\end{tabular}
\caption{Voice Assistant characteristics.}
\label{tab:VA}
\end{table*}

\subsection{Voice User Interface Devices}
\begin{figure}[t]
\centering
\includegraphics[width=3.3in]{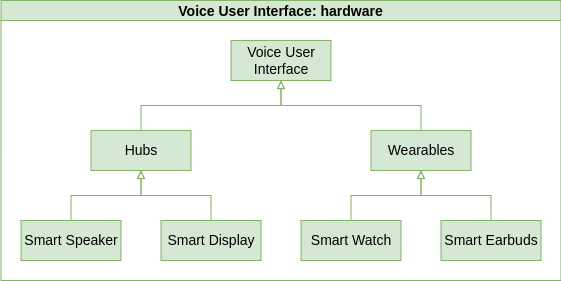}
\caption{VUI devices}
\label{fig:vuis}
\end{figure}

Voice control interfaces are divided into two categories: hubs and wearable devices. Depicted in Figure \ref{fig:vuis}, hubs are further divided into smart speaker devices and smart display devices, and wearables are further divided into smartwatches and smart glasses. We identify products that fall into each category, and that are either manufactured by, or compatible with the voice assistants of Google, Amazon and Apple to allow for comparison between these commercially-offered smart home systems. Products identified are detailed in terms of their compatibility, connectivity, processor type, input types and output mechanisms in Table \ref{tab:VUI}. From here, we can examine and compare commercial VUI products.

As with the voice assistants, findings reflect the lack of variation between market-leading VUI devices. In terms of connectivity, all use Wi-Fi and/or Bluetooth, with some products using Thread, and smart watches using NFC for contactless payments. With regard to interaction mechanisms, all products are equipped with a microphone and speaker, enabling voice interaction, and several products additionally have touch-screen displays. A variety of sensors are seen to be used to collect various types of information and expand the device capabilities, particularly on smart watches that often monitor factors of the users health. Most products are CPU-based, with exception to some small wearable devices like smart earbuds. 

In examining market-leading smart home UI devices on a functional and hardware level, we have somewhat characterised the current smart home VUI market. The taxonomy supports thinking towards an alternative set of programmable, cost effective devices that could replicate the functionality of commercial VUIs in the design of a private smart home. Additionally, the findings can be used to compare the VUI devices commercially available for smart homes, with those used in academic implementations in Section 5.

\begin{table*}[t]\footnotesize
\centering
\begin{tabular}{|p{2cm}|p{2cm}|p{2cm}|p{1.2cm}|p{6.8cm}|p{1.6cm}|}
\hline
{\bf Product name} &
  {\bf Compatibility} &
  {\bf Connectivity} &
  {\bf CPU/MCU} &
  {\bf Input types (incl. sensors)} &
  {\bf Output types} \\
\hline \hline
\multicolumn{6}{|c|}{\bf Smart Speaker: Hubs: VUI: hardware} \\
\hline 
Google Nest Mini & Google Assistant & Wi-Fi, Bluetooth & CPU & Microphone, capacitive touch, ultrasound sensing & Speaker \\
\hline
Apple HomePod Mini & Apple Siri & Wi-Fi, Bluetooth, Thread, Ultra Wideband chip & CPU & Microphone, touch control & Speaker, LED lights \\
\hline
Amazon Echo Dot (5th Gen) & Amazon Alexa & Wi-Fi, Bluetooth & CPU & Microphone, buttons & Speaker, LED lights \\
\hline
\multicolumn{6}{|c|}{\bf Smart Display: Hubs: VUI: hardware} \\
\hline
Google Nest Hub (2nd Gen) & Google Assistant & Wi-Fi, Bluetooth & CPU & Microphone, touch screen, ultrasound sensing, ambient light, motion, temperature sensors & Speaker, screen display \\
\hline
Amazon Echo Show 8 & Amazon Alexa & Wi-Fi, Bluetooth & CPU & Microphone, camera, touch screen, buttons & Speaker, screen display \\
\hline
\multicolumn{6}{|c|}{\bf Smart Watch: Wearables: VUI: hardware} \\
\hline
Samsung Galaxy Watch 4 & Google Assistant & Wi-Fi, Bluetooth, NFC, GPS & CPU & Microphone, barometer, accelerometer, gyroscope, optical heart rate sensor, electrical heart sensor, bioelectrical impedance analysis sensor, light sensor, geomagnetic sensor, hall sensor & Speaker, screen display \\
\hline
Apple Watch S3 & Apple Siri & Wi-Fi, Bluetooth, NFC, GPS & CPU & Microphone, force touch, barometric altimeter, optical heart rate, accelerometer, gyroscope, ambient light sensors & Speaker, screen display \\
\hline
\multicolumn{6}{|c|}{\bf Smart Buds: Wearables: VUI: hardware} \\
\hline
Google Pixel Buds A Series & Google Assistant & Bluetooth & MCU & Microphone, capacitive touch, motion-detecting accelerometer, IR proximity sensor & Speaker \\
\hline
Apple AirPods (3rd Gen) & Apple Siri & Bluetooth & MCU & Microphone, motion and speech detecting accelerometers, skin-detecting and force sensors & Speaker \\
\hline
Amazon Echo Buds (2nd Gen) & Amazon Alexa & Bluetooth & MCU & Microphone, accelerometer, capacitive touch, proximity sensor & Speaker \\
\hline
\end{tabular}
\caption{Voice User Interface device characteristics.}
\label{tab:VUI}
\end{table*}

\subsection{Edge Devices}
\begin{figure*}[!t]
\centering
\includegraphics[width=6.5in]{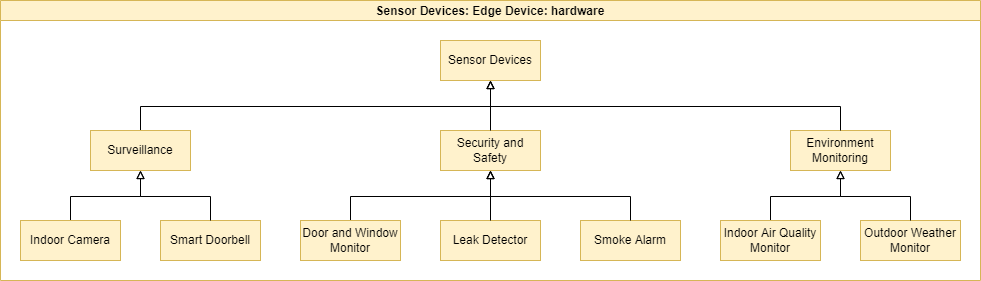}
\caption{Sensors}
\label{fig:sensors}
\end{figure*}

\begin{figure*}[!t]
\centering
\includegraphics[width=6.5in]{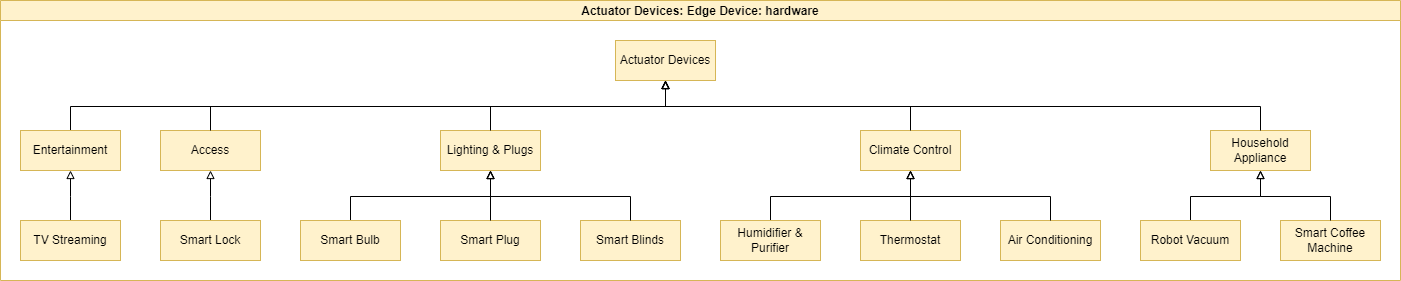}
\caption{Actuators}
\label{fig:actuators}
\end{figure*}

The two types of edge devices we have identified (sensors and actuators) are discussed in this section. We further classify sensing and actuator agents according to the type of service they enable in the smart home, which can also be seen as their perceived role in the smart home from a user perspective. For each high level category, we give examples of product types that fit within each category. The type hierarchy of sensor devices can be seen in Figure \ref{fig:sensors}, with the type hierarchy of actuator devices given in Figure \ref{fig:actuators}. 

Sensor devices available gather information for the purpose of either:
\begin{itemize}
  \item security and safety, e.g., smoke alarm
  \item surveillance, e.g, indoor camera
  \item environment monitoring, e.g., indoor air quality monitor
\end{itemize}

Actuator devices available serve to control either:
\begin{itemize}
    \item entertainment, e.g., TV streaming
    \item access, e.g., door lock
    \item lighting and plugs, e.g., smart bulb
    \item climate, e.g., air conditioning
    \item household appliances, e.g. coffee machine
\end{itemize}

For each product type, we have identified commercially available devices that are compatible with the VUI’s and voice assistants described. Each device is detailed in terms of its manufacturer, connectivity, compatibility, processor type (MCU/CPU) type, input mechanisms and output mechanisms/actions controlled. See Table \ref{tab:sensors} for a detailed list of products for each category of sensor devices. See Table \ref{tab:sensors} for a detailed list of products for each category of products under actuator devices.

The taxonomy and characterisation of edge devices helps us to understand the complexity and nature of the hardware used in smart home systems. For example, most edge devices contain MCUs, most use either Wi-Fi or Bluetooth Low Energy (BLE) for connectivity, and many interact with other home devices and phones for sending alerts, monitoring and/or control. Ultimately, there now exists a wide range of devices that are  commercially available for smart home implementations, serving a range of purposes and enabling various types of user benefit. 

We move on to examine research-based systems, in order to consider the types of device used in academic smart home implementations, where we can compare the device types used in commercial and academic smart home realms.

\begin{table*}[!t]\footnotesize
\centering
\begin{tabular}{|p{2.5cm}|p{2cm}|p{1.4cm}|p{1.2cm}|p{5.5cm}|p{2.8cm}|}
\hline
{\bf Product name} &
  {\bf Compatibility} &
  {\bf Connectivity} &
  {\bf CPU/MCU} &
  {\bf Input types (incl. sensors)} &
  {\bf Output types} \\
\hline \hline
\multicolumn{6}{|c|}{\bf Door \& Window Monitoring: Security \& Safety: Sensing Device: Edge Device: hardware} \\
\hline 
Eve Door \& Window & Apple HomeKit & BLE, Thread & MCU & Contact sensor & Notifies phone \\
\hline
Ring Alarm Contact Sensor & Amazon Alexa & Z-Wave & MCU & Contact sensor & Siren, notifies devices, phone \\
\hline
\multicolumn{6}{|c|}{\bf Smoke Alarm: Security \& Safety: Sensing Device: Edge Device: hardware} \\
\hline
Google Nest Protect (2nd Gen) & Google Assistant & Wi-Fi, BLE & MCU & Microphone, buttons, ambient light, humidity, temperature, split spectrum smoke, occupancy sensors, microphone, accelerometer & Siren, light ring, notifies phone \\
\hline
Netatmo Smart Smoke Alarm & Apple HomeKit & Wi-Fi, BLE & MCU & Photoelectric (optical) smoke sensor & Siren, notifies phone \\
\hline
\multicolumn{6}{|c|}{\bf Leak Detector: Security \& Safety: Sensing Device: Edge Device: hardware} \\
\hline
Eve Water Guard & Apple HomeKit & Bluetooth, Thread & MCU & Water sensor & Siren, light, notifies phone, devices \\
\hline
D-Link Water Leak Sensor & Google Assistant & Wi-Fi & MCU & Water sensor & Siren, notifies phone, devices \\
\hline
\multicolumn{6}{|c|}{\bf Smart Doorbell: Surveillance: Sensing Device: Edge Device: hardware} \\
\hline
Google Nest Doorbell (battery) & Google Assistant, Amazon Alexa & Wi-Fi, BLE & CPU & Camera, microphone, PIR proximity and motion sensor, magnetometer & Speaker, light ring, notifies devices \\
\hline
Ring Video Doorbell & Amazon Alexa & Wi-Fi & CPU & Camera, microphone, motion sensor & Speaker, notifies devices \\
\hline
Netatmo Smart Video Doorbell & Google Assistant, Amazon Alexa, Apple Siri & Wi-Fi & \textit{CPU} & Camera, microphone & Speaker, notifies phone \\
\hline
\multicolumn{6}{|c|}{\bf Indoor Camera: Surveillance: Sensing Device: Edge Device: hardware} \\
\hline
Google Nest Cam (battery) & Google Assistant, Amazon Alexa & Wi-Fi, BLE & CPU & Camera, microphone, motion sensor & Speaker, LED light, notifies phone \\
\hline
Ring Indoor Cam & Amazon Alexa & Wi-Fi & MCU & Camera, microphone, motion sensor & Speaker, notifies phone \\
\hline
Netatmo Smart Indoor Cam & Google Home, Amazon Alexa, Apple HomeKit & Wi-Fi & \textit{CPU} & Camera & Notifies devices \\
\hline
\multicolumn{6}{|c|}{\bf Indoor Air Quality Monitor: Environment Monitoring: Sensing Device: Edge Device: hardware} \\
\hline
Amazon Air Quality Monitor & Amazon Alexa & Wi-Fi, BLE & MCU & Temperature, CO, humidity, volatile organic compounds (VOCs), particulate matter sensors & Notifies hub, devices \\
\hline
Eve Room Indoor Air Quality Monitor & Apple Siri & BLE, Thread & MCU & Temperature, humidity, volatile organic compounds sensors & E-ink display, monitor via phone, hub device \\
\hline 
\multicolumn{6}{|c|}{\bf Outdoor Weather Monitor: Environment Monitoring: Sensing Device: Edge Device: hardware} \\
\hline
Eve Weather & Apple Siri, Eve app & BLE, Thread & MCU & Temperature, humidity, barometric pressure sensors & E-ink display, monitor via phone, hub device \\
\hline
Netatmo Weather Station & Apple HomeKit, Amazon Alexa & Wi-Fi & MCU & Temperature, humidity, barometric pressure, sound, CO2 sensors, rain gauge, anemometer & Monitor via connected devices \\
\hline 
\end{tabular}
\caption{Sensor-based product characteristics. (Note: entries in \textit{italics} are assumed)}
\label{tab:sensors}
\end{table*}

\begin{table*}[!t]\footnotesize
\centering
\begin{tabular}{|p{3cm}|p{2cm}|p{2.3cm}|p{1.2cm}|p{4.4cm}|p{2.4cm}|}
\hline
{\bf Product name} &
  {\bf Compatibility} &
  {\bf Connectivity} &
  {\bf CPU/MCU} &
  {\bf Input types (incl. sensors)} &
  {\bf Output types} \\
\hline \hline
\multicolumn{6}{|c|}{\bf Entertainment: TV Streaming: Actuator Device: Edge Device: hardware} \\
\hline 
Google Chromecast HD & Google Assistant & Wi-Fi, HDMI & CPU & via connected devices, Google voice remote & Streams media to connected TV \\
\hline
Amazon Fire TV Cube & Amazon Alexa & Wi-Fi, Bluetooth, HDMI & CPU & Microphone, buttons, via connected devices, Alexa voice remote & Speaker, LED strip, streams media to connected TV \\
\hline
Apple TV 4K & Apple Siri & Wi-Fi, Bluetooth, HDMI & CPU & via connected devices, Siri remote & Streams media to connected TV \\
\hline
\multicolumn{6}{|c|}{\bf Smart Lock: Access: Actuator Device: Edge Device: hardware} \\
\hline
Yale Smart Door Lock & Google Assistant, Amazon Alexa, Apple HomeKit & Bluetooth (Yale Wi-Fi bridge required for remote control) & \textit{MCU} & via assistants, Yale access app & Unlocks door \\
\hline
\multicolumn{6}{|c|}{\bf Smart Bulb: Lighting \& Plugs: Actuator Device: Edge Device: hardware} \\
\hline
Philips Hue Smart Bulb & Google Assistant, Amazon Alexa, Apple HomeKit & Bluetooth, Zigbee & MCU & via app, assistant & LED light \\
\hline
Nanoleaf Essentials Bulb & Google Assistant, Apple HomeKit & BLE, Thread & MCU & via app, assistant devices & LED light \\
\hline
Hey! Smart Bulb & Google Assistant, Amazon Alexa & Wi-Fi & MCU & via app, assistant devices & LED light \\
\hline
\multicolumn{6}{|c|}{\bf Smart Plug: Lighting \& Plugs: Actuator Device: Edge Device: hardware} \\
\hline
Philips Hue Smart Plug & Google Assistant, Amazon Alexa, Apple HomeKit & Bluetooth & MCU & via assistant devices, mobile app & Turn plug on/off \\
\hline
Amazon Smart Plug & Amazon Alexa & Wi-Fi & MCU & via assistant devices, mobile app & Turn plug on/off \\
\hline
Eve Energy Smart Plug & Apple Siri & Bluetooth, Thread & MCU & via assistant devices, mobile app & Turn plug on/off \\
\hline
Wemo Mini Smart Plug & Amazon Alexa, Google Assistant, Apple Siri & Wi-Fi & MCU & via assistant, mobile app & Turn plug on/off \\
\hline
\multicolumn{6}{|c|}{\bf Smart Blinds \& Curtains: Lighting \& Plugs: Actuator Device: Edge Device: hardware} \\
\hline
WEONSEE Smart Blinds Motor & Google Assistant, Amazon Alexa & Wi-Fi & \textit{MCU} & via assistant, remote, mobile app & Raise/lower blinds \\
\hline
\multicolumn{6}{|c|}{\bf Smart Humidifiers \& Purifiers: Climate Control: Actuator Device: Edge Device: hardware} \\
\hline
VOCOlinc MistFlow Smart Humidifier & Google Assistant, Amazon Alexa, Apple HomeKit & Wi-Fi & \textit{MCU} & touch control, via app, assistant, water level sensor & LED light, mist \\
\hline
VOCOlinc PureFlow Smart Air Purifier & Google Assistant, Amazon Alexa, Apple HomeKit & Wi-Fi & \textit{MCU} & touch control, via app, assistant, temperature, humidity, particulate matter sensors & LED screen, LED light, filters air \\
\hline 
\multicolumn{6}{|c|}{\bf Thermostat: Climate Control: Actuator Device: Edge Device: hardware} \\
\hline
Google Nest Thermostat E & Google Assistant & Wi-Fi, BLE & MCU & via mobile app, assistant, temperature, humidity, proximity, occupancy, ambient light sensors & LCD screen, adjust temperature \\
\hline
Netatmo Smart Thermostat & Google Assistant, Apple HomeKit, Amazon Alexa & Wi-Fi, Radio long-range & MCU & via mobile app, assistant, temperature sensor & E-paper display, temperature adjustment \\
\hline
Tado Smart Thermostat & Google Assistant, Amazon Alexa, Apple HomeKit & Wi-Fi, 6LoWPAN & MCU & via mobile app, assistant, capacitive touch buttons, temperature, humidity sensors & LED screen, adjust temperature \\
\hline 
\multicolumn{6}{|c|}{\bf Air Conditioning: Climate Control: Actuator Device: Edge Device: hardware} \\
\hline
Tado Smart AC Control V3+ & Google Assistant, Amazon Alexa, Apple HomeKit & Wi-Fi, Infrared & MCU & via mobile app, assistant, LED touch surface, temperature, humidity sensors & Controls AC unit \\
\hline
\multicolumn{6}{|c|}{\bf Robot Vacuum: Household Appliances: Actuator Device: Edge Device: hardware} \\
\hline
Samsung JetBot robot vacuum & Google Assistant, Amazon Alexa & Wi-Fi & CPU & via app, assistant, LiDAR sensor, anti-cliff sensor & Clean floors with brush \\
\hline
iRobot Roomba i7 & Google Assistant, Amazon Alexa & Wi-Fi & CPU & via mobile app, assistant, dirt detect sensor, cliff sensor, camera & Clean floors with brush \\
\hline
\multicolumn{6}{|c|}{\bf Smart Coffee Machine: Household Appliances: Actuator Device: Edge Device: hardware} \\
\hline
Smarter Smart Coffee Maker (2nd Gen) & Google Assistant, Amazon Alexa, Apple Siri & Wi-Fi & \textit{MCU} & via mobile app, assistant & Makes coffee \\
\hline
\end{tabular}
\caption{Actuator-based product characteristics. (Note: entries in \textit{italics} are assumed)}
\label{tab:actuators}
\end{table*}

\section{Smart Homes: Implementations \& their Evaluation}
This section enumerates gateway-based voice-controlled smart home implementations seen in academic literature. Focus is placed on both online and offline architectures, where we focus on the device types, control mechanisms, and evaluation methods employed. The work contributes towards providing a review of the current academic efforts in implementing and evaluating voice-control in smart home systems, so that we can assess current progress and bridge the gap between academic and commercial efforts in this realm.

\cite{Kamdar2017-uq} discuss methods used to compare and evaluate home automation systems with voice control, considering factors of flexibility, robustness, security, cost and response time. The work highlights challenges such as speech recognition in noisy environments, training voice recognition modules, the number of commands voice recognition modules can store, and the response time of the devices. We look to some more recent voice-controlled home automation systems, and consider the evaluation methods employed, as well as the challenges that arise and limitations of the systems. 

\subsection {Online Implementations}
Voice-controlled smart home systems discussed in the literature often rely on cloud processing, and we highlight such works here.

\cite{Kodali2017-uy} use an ESP8266 MCU in their smart home implementation, with the objective of implementing a cost effective, robust and scalable system. Domestic appliances, including a light, fan, bulb, and charger, are connected via relay channels to the MCU. An Android app allows for the remote control of the appliances via touch screen buttons or voice control. The system accepts simple voice commands such as “turn on” and “turn off” followed by the appliance name, and commands are recognised using Google speech to text. Experimentation results simply show the system to work as expected, using both voice and button command inputs.

\cite{Rani2017-gs} detail the implementation of an IoT-based voice-controlled home automation system similar to the previous. A mobile device is the central controller and cloud-based NLP performs command interpretation. Arduino Boards with a low power MCU and Wi-Fi connectivity are interfaced with appliances and programmed to respond to commands interpreted by the mobile device. The proof-of-concept implementation includes a fan, light, coffee machine and door alarms, however details of evaluation and the NLP software used are omitted. 

\cite{Poongothai2018-pf} employ an open-source Google Assistant to provide remote voice-control of devices in an IoT lab, where an Android smartphone application serves as the VUI and allows for both voice and text input.  Appliances include lights, fans, a projector and air conditioner, which are connected to Wi-Fi via NodeMCU devices, allowing them to be remotely controlled and monitored. Testing takes the form of checking if the sensor readings are correctly displayed, checking voice commands work (e.g. lights come on when requested), and measuring the energy consumption of devices using a current sensor. While easily extendable, the setup is very simple and evaluation methods are limited.

\cite{Putthapipat2018-qb} implement a voice-controlled smart home system, where a CPU-based Raspberry Pi is used as the central controller, with a speaker and microphone attached to allow for VUI capabilities. Voice recognition is performed using Google Cloud API, Wit.ai identifies the user's intent and the response is output via Google translate for text to speech conversion. The system requires a stable internet connection to work, the energy consumption is high, the speech recognition performance depends on the quality of the microphone and there is no built-in screen. The advantages of the design lie in its flexibility and scalability, and the potential for more hardware devices to be added. 

\cite{Kumar2021-ow} use a mobile device interfaced with Google's DialogFlow API to allow voice-control of appliances in a smart home system. A Raspberry Pi serves as the central server, and ESP8266 and Arduino Mega 2560 MCUs are used to allow remote control home appliances. Additional IR sensors are used for object detection to turn lights on/off and PIR sensors are used to turn on/off fans. The experimental setup is proposed, but there are no evaluation methods employed.

\cite{Sudharsan2022-yf} propose a prototype Alexa smart speaker composed of a Raspberry Pi, ReSpeaker v2, Raspberry Pi camera v2 and a regular speaker. OpenCV provides a face recognition algorithm which is used to enable user authentication, and the Alexa voice service SDK is used for voice interaction. Snowboy wake-word engine is employed to prevent accidental activation of the smart speaker, and improve privacy, where the accuracy of this function is measured by the false alarm per hour vs the miss detection rates. The evaluation methods are not specifically described, however observations are made, e.g. there is a 0.5 second delay after wake word detection, after which audio is sent to the Alexa cloud service.

\subsection {Offline Implementations}
We turn now to look at literature on offline implementations of voice-controlled smart home systems. Offline is used in the sense of not using third party, typically cloud-based, services, rather than meaning the system is disconnected from any network.

\cite{Arriany2016-cv} use Windows 7 speech recognition software for voice control of a smart home fan and light. A laptop with Windows 7 OS serves as the UI, a PC acts as the home server, and an Arduino UNO MCU is used to process commands sent from the home server and route teach to the intended device. For evaluation, the word recognition accuracy is measured in noisy and quiet environments, and when using varying quality microphones. Execution time and general system functionality are additionally tested. The authors suggest the use of more specific means of testing in future work, where data rate and error rate could be measured.

\cite{Ali2017-am} implement voice-control of electrical devices in the home and office, using the EasyVR 2.0 \footnote{https://www.sparkfun.com/products/retired/12656} voice recognition module to handle command inputs. An MCU connects to appliances via relays, and RF communication is used to transmit commands from the voice recognition module to the MCU. Experimentation is performed by repeating each command 30 times. The success rate of command recognition for different age groups and genders is measured, as well as when different types of noise or physical obstacles are present. Noise and obstacles were seen to most negatively impact recognition performance. 

Again the EasyVR Shield 2.0 is used for voice-control of appliances connected via relays to an Arduino MCU in \cite{Elsokah2020-ur}. The work implements a voice-controlled smart room, where the lighting, radio, television, music player and air-conditioning can be controlled using voice commands. Testing took the form of repeating spoken commands in different noise and weather conditions, with participants of different genders and ages. The average success rate of command recognition was seen to be 96\%. The authors also considered the price and quality of system components, the response speed, and the overall system quality. Future work includes using a Raspberry Pi the provide the system greater AI capabilities, and using mobile phone input to increase the user-friendliness.

\cite{Ehikhamenle2017-er} implement a wireless voice recognition system to detect a finite set of commands, using Arduino Uno microcontrollers, a relay circuit, an Arduino v3 voice recognition module, plus an RF transmitter and receiver. When a command is recognised, the MCU coordinates the execution of the command via the relay. The voice input device consists of a microphone, ultrasonic sensor, and the voice recognition module. The v3 voice recognition module was trained on a single user, and speech recognition performance was tested by placing the user and device in a quiet and noisier room, finding music to interfere with recognition accuracy. Due to the v3 module being trained on a single-person, the command recognition system is speaker-dependent, and therefore more errors occurred when tested on a different user.

\cite{Munir2019-md} detail the development of a speaker-dependent offline smart home system, where speech recognition is also performed on an Arduino v3 module. ESP8266 devices are used to enable wireless communication, and the OpenCV library is used on a Raspberry Pi for face recognition to allow for automatic access control (door opening). The v3 module performs speech to text and text to speech functions. The accuracy of the speech and face recognition models are reported to be 90\% and 96\% respectively with low latency, but the specific evaluation methods are not described. The v3 module limits the system since it can only store 80 commands and is trained on a single person.

Similarly, \cite{Shehab2020-hx} implement a voice and gesture controlled home automation system that uses an Arduino v3 voice recognition module to take speech as input. An Arduino Mega 2560 MCU acts as the central controller, connecting to a light, alarm, fan, and air conditioning via a relay. A display shows the status of devices, and there an ultrasonic sensor allows for the detection of four hand gesture patterns. Testing saw the system used by different types of patients, and the success rate of voice and gesture interaction were measured.

Alternatively, \cite{Bhagath2021-gk} use the open-source PocketSphinx library to develop an offline embedded speech recognition system for Android-based mobile devices that can be used in a home automation system. A Raspberry Pi 3 serves as the smart home controller in the prototype implementation, with LED lights connected. When tested with live speech the system saw 80\% accuracy, with a delay time of 1 second for command recognition and execution. 

\cite{Bai2022-hl} detail the design of a speech recognition algorithm for a voice-controlled smart home. The algorithm uses feature extraction to extract speech features, and a template library for pattern matching, where the template with the highest similarity to the speech features is returned as the recognition result. For evaluation, voice files containing recordings of commands were played to the system, e.g. “Help” was repeated 20 times and was recognised with a 95\% accuracy rate, “Fire extinguishing” was repeated 20 times and was recognised with a 85\% accuracy rate. The average speech recognition accuracy reported is 83.75\%, where the authors identify microphone quality and the clarity of pronunciation tn negatively impact performance.

\section{Voice Control: Methods, Tools, \& Evaluations}
There is much interest within the research community for the development of an offline speech recognition system that could help towards the creation of a cloudless voice assistant \cite{Murshed2021-ng}. In this section, we consider available libraries and tools for implementing voice assistant, ASR and KWS functionalities on resource-constrained devices. The methods and metrics used to compare and evaluate these components are also of interest, as we look towards potential tools for implementing an on-device voice assistant, as well as an evaluation framework for measuring the performance of this. 

\subsection{Voice Assistants}
\cite{Eric2017-il} compare voice assistants that could be integrated with an embedded home automation system, including Jasper, Google Cloud Speech API, Alexa Voice Service, and Bing Speech API. The characteristics of each are compared, considering programming languages, supported human languages, supported architectures and whether each is open source and works offline or online, e.g. Jasper is open source, modular in design, programmed in Python, and supports an offline voice enabled gateway architecture (when using offline STT and TTS engines). The study provides some insight into available offline voice assistant systems. 

Rhasspy \footnote{https://rhasspy.readthedocs.io/en/latest/} provides an open source, offline set of voice assistant services, and has seen implementation in a private Raspberry Pi-based smart home voice assistant \cite{Dallmer-Zerbe2021-lp}. The implemented system saw low response times, however transcription accuracy is sub-optimal and would need to be improved, perhaps through employing a more sophisticated language model and a microphone that filters noise. In a comparison of voice assistants, Rhasspy was seen to perform best in terms of trustworthiness, security of transmission, and on-device intelligence, where Alexa, Google Assistant and Mycroft were also considered \cite{Jesse2021-ty}. 

Voice Assistants can generally be assessed by checking if they perform the desired functionality. For example, scenarios are given to evaluate the functionality of the smart home dialog system in \cite{Huang2015-db}, and \cite{Chen2018-ei} evaluate their voice assistant by measuring the intent recognition accuracy and the entity recognition accuracy for the functions available (e.g. dialing a contact, checking the weather in a location).

\subsection{Automatic Speech Recognition}
Open-source speech recognition models for edge devices are compared in \cite{Peinl2020-fv}, providing useful methods for comparing such systems. Each ASR model is run on Raspberry Pi 3 and Nvidia Jetson devices, and metrics of real-time factor (RTF) and accuracy (WER) are measured using the LibriSpeech dataset \cite{Panayotov2015-uu}. Amongst Mozilla DeepSpeech \footnote{https://github.com/mozilla/DeepSpeech}, and Facebook wav2letter \footnote{https://github.com/facebookresearch/wav2letter}, PyTorch Kaldi \footnote{https://github.com/mravanelli/pytorch-kaldi} was found to be the most effective model. Kaldi is widely used, and \cite{Pinto2020-sc} implement a Kaldi ASR system on a mobile device containing an ARM CPU and a low-power GPU. It is also notable that a custom version of a Kaldi training recipe was used in the design of the (now unavailable) Snips Voice Platform for SLU, where LibriSpeech was again used for model evaluation, and performance metrics included WER and speech and memory usage \cite{Coucke2018-ay}. 

Using similar methods to those above, the transformer-based speech recognition systems Wav2Vec 2.0 and Speech2Text are compared by running each on Raspberry Pi and Nvidia Jetson Nano devices in \cite{Gondi2021-jo}. The LibriSpeech dataset \cite{Panayotov2015-uu} is used for testing, and each models performance is measured in terms of latency, accuracy and computational efficiency (CPU and memory footprint).

A dataset of one speaker recorded at varying distances away from the microphone has been used to evaluate the WER of the lightweight, open source voice recognition systems, Julius and PocketSphinx, running on a Raspberry Pi 3 \cite{Vojtas2018-jk}. Julius was found to perform better in terms of word recognition probability, and the evaluation method is useful to see since the maximum effective distance for speech recognition can vary significantly between platforms \cite{Heartfield2018-ne}. PocketSphinx is also employed for speech recognition functionality in the implementation of a cloud-free local voice assistant in \cite{Polyakov2018-ff}.

We consider some additional libraries for ASR, and their evaluation. Tensorflow Keras Library is used to implement a deep learning model for command classification on  Raspberry Pi 3 in \cite{Zonios2021-ry}. The classification model achieved 87.8\% accuracy and 1.136 second latency on an 8 command recognition task, where k-fold cross validation was used for evaluation. The test dataset contained spectrograms for 905 voice samples, each 4 seconds long and recorded by a male and female on both headset and mobile phone microphones. Additionally, an open-source end-to-end ASR toolkit named ESPnet has been developed and seen to achieve reasonable performance on WSJ, CSJ and HKUST ASR tasks \cite{Watanabe2018-ta}. In recent times, end-to-end approaches are generally proving to achieve better results in a range of speech processing tasks, compared with conventional pipelines.

\subsection{Keyword Spotting}
Keyword spotting (KWS) is an always-on feature that often serves as the entry point for speech-based smart home devices. The problem involves detecting a predefined command from a continuous stream of audio, and the always-on nature means KWS models are normally implemented on very small MCUs  \cite{Fernandez-Marques2018-al}. For example, Snowboy hotword-detection is an offline model that can detect one particular word to wake a system \cite{Amberkar2018-nf}.

Accuracy for commercial keyword spotting algorithms is high. EdgeSpeechNets \cite{Lin2018-xx} and TinySpeech \cite{Wong2020-vw} are two speech recognition systems that were designed to run on edge devices, and evaluated using the Google Speech Commands dataset \cite{Warden2018-ja}. The dataset was designed to train and evaluate keyword spotting systems, and contains 105,829 utterances of 35 words, with each utterance stored as a one-second WAVE format file. EdgeSpeechNets best run achieved ~97\% accuracy on the test set, and TinySpeech best model achieved ~95\% accuracy. More recently, a transformer-based architecture set a benchmark for the Google Speech Commands dataset, achieving 98.6\% and 97.7\% accuracy on the 12 and 35-command tasks respectively \cite{Berg2021-al}.

Metrics relating to accuracy, model size in bits, and power usage to evaluate their keyword spotting algorithm in \cite{Blouw2020-sb}. The network is trained using a train/dev/test split dataset containing one second speech samples belonging to twelve command classes, and the accuracy of various sized models are compared to gauge the trade-off between recognition accuracy and model size. In a similar way, \cite{Zhang2017-vn} compare neural network-based KWS models in terms of accuracy and memory requirement . A depthwise-separable convolution neural network (DS-CNN) was found to be best, achieving ~94\% accuracy and requiring 38.6KB memory.

Along with accuracy, the metrics false accept (FA) rate and false reject (FR) rate are commonly used to evaluate keyword spotting systems, where lower rates result in better user experience \cite{Michaely2017-ss}. A positive dataset containing utterances that begin with the trigger phrase can be used for measuring WER and FR rate, and a negative dataset containing utterances that were accepted by a trigger-phrase detector but do not contain the trigger phrase can be used to measure the FA rate. \cite{Myer2018-xo} use the Google Speech Commands dataset \cite{Warden2018-ja} to evaluate a KWS model, where accuracy, FA rates and FR rates are measured in both clean and noisy environments. Plus, an additional dataset is used to test the system in a wider variety of acoustic conditions.

\section{Towards a more ideal smart home}
We look towards the design of a prototype smart home system that can mirror the benefits provided by commercial smart homes, while ensuring privacy and security. \cite{Kamdar2017-uq} suggest an ideal smart home balances cost, robustness, reaction time, processing power, flexibility, security and convenience, therefore we use these factors to imagine a better voice-controlled smart home system that is respectful of user's needs. The section details technologies, devices and methods that can help towards building a better voice-controlled smart home.

\subsection{Voice-based Authentication}
Most security problems are seen to be related to the lack of authentication schemes for users and devices \cite{Alam2012-zl}. Lack of user authentication can allow for  hidden voice attacks, where ultrasonic voice commands and voice commands unintelligible to humans have been found to be recognised by the majority of commercial voice controlled systems \cite{Heartfield2018-ne}. In one case, a Google system was able to recognise the phrase “Ok Google” with 95\% accuracy, compared to human transcriber’s 22\%. The absence of user authentication allows anyone in the home to interact with and extract information from your smart home, including TVs which can wake and interact with devices. An overview of cybersecurity risks that arise from the absence of authentication schemes in smart homes are described in \cite{Sudharsan2022-yf}. We consider voice-based methods for implementing secure user authentication in smart homes. 

Voice-based biometric authentication is advantageous in that it is part of the user,  and the authentication process can be done hands-free. Despite the benefits, voice authentication is not as accurate or secure as other biometric methods, e.g. voice clips of a speaker can be used to impersonate the speaker (replay-attacks) \cite{Mukhopadhyay2015-uf}. In response to the weakness of purely voice-based user authentication, a continuous speaker authentication system has been implemented \cite{Feng2017-vk}. A microphone is used in combination with an accelerometer on a wearable device to check the speech received by the VUI originated from the speaker's throat. The system achieved an average overall detection accuracy rate of 97\%, with a 0.09\% false positive rate, and appears to be an effective solution.

\subsection{Smart Home Usage}
User-centric design should be looked towards in the design of an ideal smart home. One way to achieve this is to gauge user preferences, and find how existing smart home users interact with their systems and the benefits that are most popular. For example, one study found media devices (smart TVS and streaming devices) to be the most common type of smart home device in the majority of world regions, with surveillance devices being the most common in South and Southeast Asia \cite{Kumar2019-mq}.

In terms of user queries, an analysis of smart speaker voice history logs has shown music to be the most common query, followed by information \cite{Bentley2018-hm}. Users generally made use of three domains on a weekly basis, the majority of commands were single sentence commands, and stop was the most common command made to Google Home. Another study on interactions with smart speaker devices found that requests for audio were most common, followed by requests to control media and requests to control smart home devices \cite{Malkin2019-qg}.

Human evaluation can be used to gauge the strengths and weaknesses of a smart home system. The Sweet-Home voice-based system has been used for evaluating a voice-controlled smart home for seniors and people with visual impairment to see if the voice assistant capabilities are effective enough for use in real-world scenarios \cite{Vacher2015-dv}. The paper explores qualitative and quantitative methods of evaluation, and findings showed that some people did not like the rigid grammar, instead preferring more natural commands. The absence of user feedback in the system (to respond to user commands) was noted to negatively impact user experience as it was unknown whether the system had interpreted the given command. The experiments were conducted using scenarios to demonstrate the user performing different tasks using the system. There are always trade-offs to be made in ASR systems, for example it is easier to transcribe utterances if a strict command syntax is imposed, however users often deviate from strict grammars \cite{Mishakova2019-qd}.

\subsection{Model Personalisation}
Following on from our discussion on smart home usage, we consider the possibility to personalise speech recognition models to users, and in turn improve the robustness and performance. It is predicted that speech recognition models of the future will be greatly personalised to individual devices, through on-device methods of training lightweight models \cite{Hannun2021-dn}. In \cite{Gashi2022-it}, personalised models are shown to perform significantly better than population models for sleep quality recognition and sleep stage detection using wearable devices. Additionally, narrower age-dependent modelling produced higher depression detection accuracy, when compared with age-agnostic modelling. Gender-dependent systems are also commonly used for improving accuracy \cite{Stasak2022-wu}.

Personalised speech recognition has been used to help handle individuals' distinct perception of physical objects in their house in \cite{Mehrabani2015-as}. For example, a window in the living room can be called “living room window”, “first floor window" or “the big window” by different users. The system allows users to select customised names for their devices which are integrated into the language mode. It is thought that such customisable communication with devices allows for a more natural interaction between humans and the IoT. Similarly, \cite{Rubio-Drosdov2017-qo} recognise that constrained sets of commands can be troublesome for voice interface systems, therefore a system that associates actions with multiple descriptive tags has been developed.

Google found integrating personal information into the language model of a large vocabulary speech recognition system for mobile devices to be advantageous for reducing the WER without increasing computational overhead, e.g. incorporating the users list of contact names to reduce the number of out-of-vocabulary words \cite{McGraw2016-ow}. Personal Voice Activity Detection (VAD) has also been published by Google to help achieve state-of-the-art performance in an on-device speech recognition system \cite{Ding2022-cf}. Personal VAD is an always-running component that detects the voice activity of given target speakers, and can help in improving speech recognition and reducing computational resources.

Relating to the field of model personalisation, on-device training paradigms for training and fine-tuning models are increasing in popularity, due to users becoming more privacy aware, and legislation protecting users against the storage of their data \cite{Almeida2021-mf}. A framework for personalising convolutional neural networks (CNNs) using on-device resources, that makes use of early exits to improve efficiency, and that could be applied to speech recognition scenarios has been published \cite{Leontiadis2021-ve}.

\subsection{TinyML \& Edge Computing}
TinyML aims to bring machine learning (ML) inference to low-power IoT devices, and the field has seen some growth in recent years. The field is highly relevant to the paper, in that we are examining the feasibility of implementing TinyML algorithms for speech recognition on edge devices in the home. Several software stacks have been released to implement and train TinyML algorithms, and TensorFlow Micro has been used for benchmarking and analysing the onboard performance of 30 neural network (NN) models running on 7 available MCU boards in \cite{Sudharsan2021-pz}. Metrics of price-performance ration, onboard accuracy, and memory consumption are used to compare the results. 

Similar works exist. \cite{Zhang2018-kq} compare the performance of ML packages for edge devices, where metrics of latency, memory footprint and energy usage are considered. Hardware platforms for edge computing are discussed in \cite{Hadidi2019-ws}, and popular frameworks for edge inference are also compared in terms of execution time, energy consumption and temperature. An Open Framework for Edge Intelligence (OpenEI) has been presented in \cite{Zhang2019-sx}. OpenEI has been designed to address challenges in edge computing, relating to computing power mismatches between existing AI algorithms and edge platforms, and the difficulty of data sharing between edge devices. The framework can support multiple applications, including smart homes.

\subsection{DIY Technologies \& Devices}
We can use the taxonomy of smart home interface devices to imagine a similar collection of devices that can proxy the commercial products we have detailed. In a previous study of commercial products, \cite{Koreshoff2013-eg} highlight the potential for researchers to rethink the approaches used, and use DIY technologies to quickly and cheaply construct prototype systems to explore the vision of the IoT.

Arduino and Raspberry Pi devices commonly appear in literature surrounding such smart home implementations, and a comparison of systems built using Raspberry Pi and Arduino controllers can be seen in \cite{Gunge2016-rc}. Low power, low-cost sensors for various smart home applications that work with Arduino and Raspberry Pi devices are compared in \cite{Gazis2021-gk}. Single-purpose units designed for indoor air quality monitoring have also been compared \cite{Omidvarborna2021-cn}, and the technical specifications of low-cost sensors for measuring air quality are collated in \cite{Demanega2021-xb}. The works are useful to refer to when designing a research-based smart home IoT system, where the advantages and limitations of these  available DIY technologies could be explored.

\subsection{Open-Source Datasets}
Datasets are vital for evaluation of a system, and can also be used for training purposes. Available datasets that can be used for both tasks include:

\begin{itemize}
    \item Mozilla’s Common Voice dataset \footnote{https://commonvoice.mozilla.org/en} contains over 2k hours of voice recordings from over 80k English speakers between the ages of 19 and 79. Each entry consists of an MP3 recording and the corresponding text file transcription. 
    \item LibriSpeech \cite{Panayotov2015-uu} is a corpus containing 1000 hours of read English speech. The labels are assigned at a sentence level, therefore there is limited word-level alignment. LibriSpeech is therefore more suitable for automatic speech recognition, rather than keyword spotting.
    \item Amazon has released an open-source speech dataset with the aim of encouraging developers to build more third-party apps and services for its smart speaker Alexa. The dataset contains one million spoken samples across 51 languages. Open-source code has been released to help developers train multilingual AI models \cite{Quach2022-jp}.
\end{itemize}

\section{Conclusion}
The paper arises from an investigation into the feasibility of implementing a voice-controlled smart home system that processes audio data locally on IoT devices, and overcomes some consumer issues of control, privacy, security, power usage and cost. We have examined hardware and software components of a smart home in our taxonomy, and reviewed academic voice-controlled smart home implementations and their evaluation. Additionally, we have considered available libraries for implementing voice control functionalities on resource-constrained devices, and identified potential avenues to explore in continuing research towards privacy-preserving voice control for smart homes. The taxonomy establishes clear terminology surrounding IoT-based smart homes, and provides insight into the nature of various commercial smart home devices. The approach taken tries to give a holistic reflection of current technologies used in voice-controlled smart homes, taking into account commercial and academic efforts and identifying key areas of improvement. Overall, the work hopes to support research towards methods for developing and evaluating a voice-controlled smart home system that processes speech at the edge, and is private and secure by design.


\bibliographystyle{apalike}
\bibliography{Bibliography}

\vfill
\end{document}